# Powerful terahertz emission from $Bi_2Sr_2CaCu_2O_{8+\delta}$ mesa arrays

T. M. Benseman[1], K. E. Gray[1], A. E. Koshelev[1], W.-K. Kwok[1], U. Welp[1], H. Minami[2], K. Kadowaki[2], and T. Yamamoto[3]

[1]Materials Science Division, Argonne National Laboratory, Argonne, Illinois 60439, USA

[2]Institute for Materials Science, University of Tsukuba, Ibaraki 305-8753, Japan

[3]Wide Bandgap Materials Group, Environment and Energy Materials Division, National Institute for Materials Science, 1-1 Namiki, Tsukuba, Ibaraki, 305-0044, Japan

Stacks of intrinsic Josephson junctions (IJJs) in high-temperature superconductors enable the fabrication of compact sources of coherent terahertz radiation. Here we demonstrate that multiple stacks patterned on the same $Bi_2Sr_2CaCu_2O_{8+\delta}$ crystal can - under optimized conditions - be synchronized to emit high-power THz-radiation. For three synchronized stacks we achieved 610 microwatts of continuous-wave coherent radiation power at 0.51 THz. We suggest that synchronization is promoted by THz-waves in the base crystal. We note though that synchronization cannot be achieved in all samples. However, even in these cases powers on the 100-μW scale can be generated.

There is rapidly growing interest in the generation of electromagnetic (EM) waves at terahertz frequencies (1 THz = $10^{12}$ c/sec), because of their potential applications in novel nondestructive imaging and spectroscopy in a wide range of settings. These include not only the physical, chemical, and biological sciences, but also pharmaceuticals, manufacturing, environmental monitoring, medical diagnostics, high-bandwidth communication technologies, security applications, and defense purposes [1]. For these applications, it is highly desirable to have sources of THz radiation that are compact and stable, with power levels in the milliwatt range. A variety of technologies have been developed [1, 2]; nevertheless, the frequency range from 0.5 to 1.3 THz – the so-called THz-gap – has been difficult to fill with solid-state sources.

Recently, we have demonstrated that stacks of intrinsic Josephson junctions in the highly anisotropic high-$T_c$ superconductor $Bi_2Sr_2CaCu_2O_{8+\delta}$ (Bi-2212) can be induced to emit coherent continuous-wave radiation in this frequency range [3]. These samples were designed in such a way that an electromagnetic cavity resonance promotes synchronization of a large number of intrinsic Josephson junctions [4] into a macroscopic coherent state. Power levels of up to 80 µW [5] from a stand-alone mesa and frequencies up to 1 THz [6] have been reported, while Orita *et al.* have demonstrated synchronized emission from two mesas on the same crystal [8]. Nonetheless, power levels reported to date have been significantly below what is required for practical applications, and what has been theoretically predicted to be possible [9]. Here we describe the generation of power levels as high as 0.6 mW of continuous-wave radiation from compact solid state sources in the THz gap frequency range.

When a DC voltage $V_J$ is applied across a Josephson junction, high-frequency electromagnetic oscillations will be generated at the Josephson frequency $f_J = V_J/\Phi_0$, that is, 1 mV per junction corresponds to 0.482 THz. Here, $\Phi_0$ is the flux quantum. For a long rectangular mesa-shaped sample, emission occurs when the Josephson frequency is close to the cavity resonance, whose frequency is given by $f = c(T)/2w$. Here $c(T)$ is the temperature-dependent far-infrared light speed in the Bi-2212 mesa [10, 11], while $w$ is the mesa's width.

Achieving significant emission power depends on phase synchronization of the junctions in the stack, and it was found in [3] that the radiated power was proportional to the square of the number of junctions switched to the resistive state, implying that at the correct bias voltage junctions emit coherently. At the resonance condition, resistive power dissipation in the mesa is of the order of tens of mW. Consequently, one limit on the feasible number of junctions in a stack – and thus the THz power it can generate – is removal of heat through the base of the mesa [12]. Lithographic process constraints also limit the mesa height to a few microns at most. For both of these reasons, it is highly advantageous if many mesas can be made to radiate coherently. The THz power would then scale as the square of the number of mesas as long as all mesas fall within one free-space wavelength at the cavity resonance frequency (i.e. up to three or four mesas) and would then scale linearly as the number of mesas thereafter. Phase synchronization between adjacent mesas may be achieved via electromagnetic coupling through the bulk of the crystal, as shown in the top inset of Fig. 1. Part of the THz power generated in individual mesas escapes as leakage radiation into the base crystal [13, 14]. Since the leakage radiation contains mostly the same in-plane wavevector as the mesas, i.e., $\pi/w$, efficient synchronized coupling is expected for mesas spaced at a distance of $w$.

Here we describe the results on two mesa arrays, both fabricated on optimally doped Bi-2212 crystals. The first consists of 8 parallel $350 \times 60 \times 0.75$ μm$^3$ mesas fabricated using optical lithography and argon ion milling on a crystal that has been mounted on a sapphire substrate using silver conductive epoxy. The second contains 6 parallel $400 \times 60 \times 1.1$ μm$^3$ mesas on a crystal that has been soldered to a Cu substrate. A device of this type is shown in the bottom inset of Figure 1. The devices were mounted in a liquid helium flow cryostat with optical windows, and wired such that the bias current through each mesa could be controlled independently. The detection optics are configured to allow the overall intensity and power spectrum of the radiation to be measured simultaneously using silicon bolometers and a Bruker Vertex 80v FTIR spectrometer. (See Supplemental Material.)

Fig. 1 shows the temperature dependence of the resistance of three mesas in the first array. There is some spread of the value of $T_c$ and variations in the contact resistance across the array, which may affect the mutual synchronization of the mesas (see below). Initially, the bias current through an individual mesa was swept up until all junctions were driven into the resistive state, whereupon the IV-characteristics and THz-emission power are recorded on decreasing the bias current (Fig. 2). As can be seen in the plot, the IV-characteristics are strongly influenced by self-heating, to the extent that it is S-shaped, which is typical for mesas of this size [12, 15]. For the mesas on this chip, THz emission occurs where dI/dV is very large, or even negative, and it has been suggested [16] that this may be essential for some modes of THz emission in these devices. The jumps in the IV-curves around 40 mA may be a signature of the formation of hot-spots in the mesas [16]. The over-all features of the IV-curves of different mesas are very similar; however, details vary which will affect their synchronization. Maximum THz power as a function of bias voltage typically peaked around 120 µW, at a bath temperature of 40 – 55K (Fig. 2). When two or three mesas on the chip are biased, the maximum achievable THz power scales almost as the square of the number of energized mesas (see inset in Fig. 3a), up to a maximum of 610 µW for three mesas. In this case, we find that maximal emission occurs at a slightly higher mesa temperature than for one or two mesas, and at correspondingly lower bias voltage (as seen in Fig. 3a) and emission frequency (Fig. 3b). For four or more mesas on this chip, the total power dissipation becomes excessive, and it is no longer possible to cool the mesas to their optimal operating temperature.

We also find that the combined THz emission spectrum is monochromatic at 0.51 THz, to within the 2.25 GHz resolution of the spectrometer (see Fig. 3b), agreeing closely with the Josephson relation $f_J = V_{mesa}/N\Phi_0$. All of these results imply that at the correct temperature and bias settings, strongly enhanced coherent THz emission from multiple mesas is being generated. Achieving monochromatic emission may in some cases require individual adjustment of the bias currents, depending on the bath temperature and the combination of mesas used, since the heat sinking properties, contact resistances and details of the IV-characteristics of the mesas may not be identical. Consequently, the maximum radiated power from an array of mesas represents the best compromise across the properties of the varying individual stacks, with the result that the power scales as slightly less than the square of the number of stacks (Fig. 3a, inset).

The line intensity measured by our spectrometer in Fig. 3b does not scale exactly as the radiation power plotted in Fig. 3a, since the angular distribution of emitted radiation changes when multiple mesas are radiating in phase as shown in Fig. 3c. Regions "a" and "b" respectively mark the acceptance angles for the spectrometer and intensity bolometer. Whereas the emission pattern of a single mesa is similar to previous observations [7, 17], the emission for three synchronized mesas is enhanced and becomes asymmetric. The cause of this behavior is not established yet, but may result from THz voltage phase shifts between the mesas. Sufficiently large arrays of mesas form a grating on the BSCCO crystal, and ideally should enhance the out-coupling of the in-plane THz-waves into the perpendicular direction. Enhanced radiation power due to grating coupling has been demonstrated in quantum cascade lasers [18]. Since the height of the individual mesas (~0.75µm) is much smaller the THz free-space wavelength, a single mesa emits approximately spherical or cylindrical waves (see Fig. 3c). In contrast, a sufficiently large coherent array will generate a parallel beam, which is desirable for many applications.

Figure 4a shows the bias dependence of the THz emission intensities from the second array. With increasing number of activated mesas the emission power increases reaching 250 microwatts for biasing five mesas simultaneously. For each combination of mesas the array temperature has been adjusted to achieve highest power as reflected in the figure legend. Nevertheless, the total power scales approximately linearly with the number of mesas energized, rather than quadratically (see inset in Fig. 4a), indicating incoherent superposition of the emission of individual mesas. We were not able to synchronize the emission from the mesas in this array as is evidenced by the spectra shown in Fig. 4b. In all conditions, the emission line is broad and/or multi-peaked. At this time, the exact mechanism for synchronization of multiple stacks via the base crystal – and the physical conditions necessary for it to occur – are not yet fully understood. However, as shown for mesa "g" in Fig. 4a, the IV-curve for the individual mesas change as other mesas are activated. From the foregoing it appears that details of the temperature distribution and self-heating at the resonance point play an important role.

In making an accurate determination of the THz power radiated by the devices, it is essential to have a reliable optical responsivity calibration for the detectors used. Few directly calibrated radiation sources exist in this frequency range. We have calibrated the two silicon bolometers employed in this experiment at a number of wavelengths between 0.5 and 12 THz, using a reference source comprised of a blackbody cavity standard and gold mesh bandpass filters. The details of the responsivity calibration are given in the Supplemental Material. We find a strongly wavelength and device dependent optical responsivity of $6.9 \times 10^6$ V/W and $4.4 \times 10^6$ V/W at 500 μm, respectively, even though the manufacturer's quoted electrical responsivities are almost the same for both devices. This variability between sensors may also account for varying results obtained in different labs.

In summary we have generated continuous-wave coherent THz radiation with power up to 0.6 mW at 0.51 THz by simultaneously energizing multiple Bi-2212 mesas fabricated on the same single crystal. The emitted radiation appears to be monochromatic, and the total power level scales approximately as the square of the number of mesas energized, suggesting that the mesas are emitting coherently with respect to each other. We find that in practice self-heating of the mesas limits the power output and in fact may prevent synchronization of multiple mesas. On a second array, the incoherent superposition of the emission from five mesas yields an emission power of 250 μW. A better understanding is therefore needed of the relationship between device geometry, self-heating, and coupling of THz radiation between mesas.

Work at Argonne National Laboratory was funded by the Department of Energy, Office of Basic Energy Sciences, under Contract No. DE-AC02-06CH11357, which also funds Argonne's Center for Nanoscale Materials (CNM) where the patterning of the BSCCO mesas was performed. We thank R. Divan and L. Ocola for their help with sample fabrication.

Figure captions:

Fig. 1 Temperature dependence of the resistance of three mesas in array 1. Lower inset: Optical micrograph of array of 400 × 60 micron mesas. Top inset: Numerical simulation of the c-axis component of THz-frequency electric field for an array of mesas with width and spacing 60 microns, showing the electromagnetic coupling of the mesas though leakage of radiation into the Bi-2212 base crystal. (See reference [14] for details of simulation methodology.)

Fig. 2 Current-voltage characteristic and THz-emission power of three mesas in array 1 at 55K.

Fig. 3 (a) Radiation power versus bias voltage across mesa e at optimized bath temperature for one, two, and three mesas. Curve for one mesa is same as in Figure 2, albeit with intensity plotted against voltage instead of current. When multiple mesas are biased, the bath temperature must be reduced to offset the increased power dissipation and maintain the mesas at their optimal temperature. Inset shows scaling of maximum THz signal as approximate square of number of mesas.
(b) Spectra at maximum THz power (for optimized bath temperature) for one, two, and three mesas. Spectrometer resolution is 0.075cm$^{-1}$, or 2.25 GHz.
(c) Angular dependence of the emission from a single mesa and from three synchronized mesas. Regions "a" and "b" mark the acceptance angles for the spectrometer and bolometer, respectively. The data were taken with a slit 3 mm wide, corresponding to 14 degrees.

Fig. 4  (a) Bias dependence of the THz emission power of the second array for various activated mesas and IV-curves of mesa "g", (b) Emission spectrum of the second array for various activated mesas. A monochromatic emission line could not be achieved.

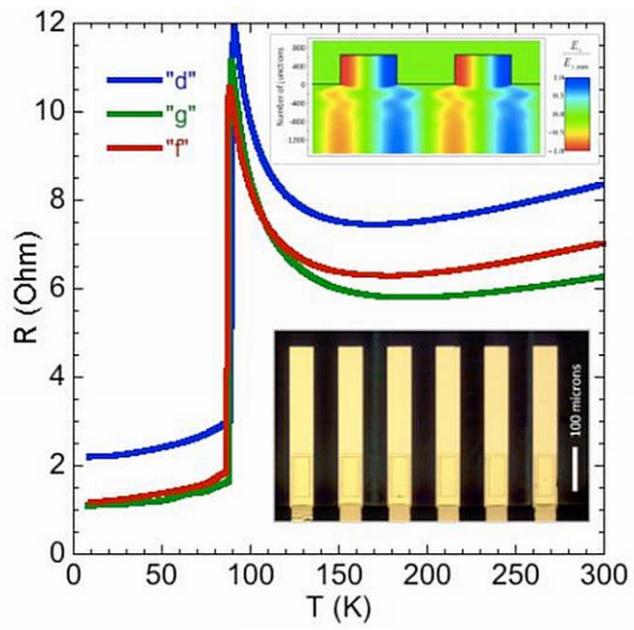

Figure 1

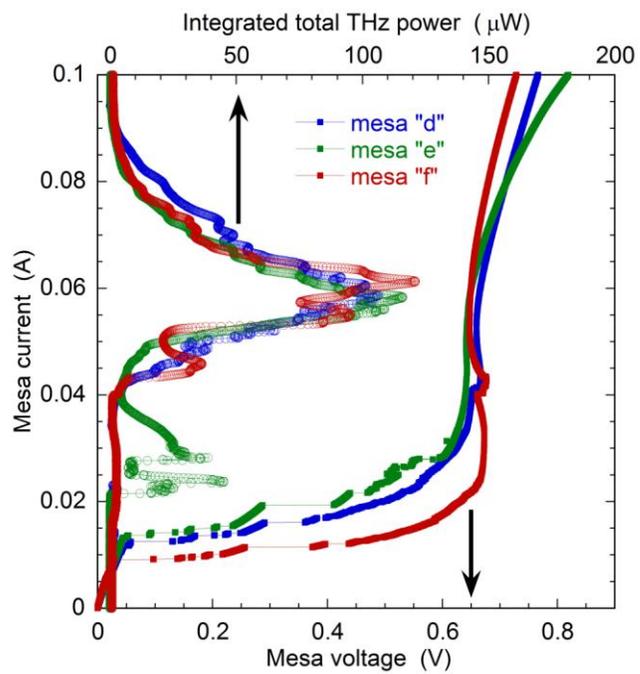

Figure 2

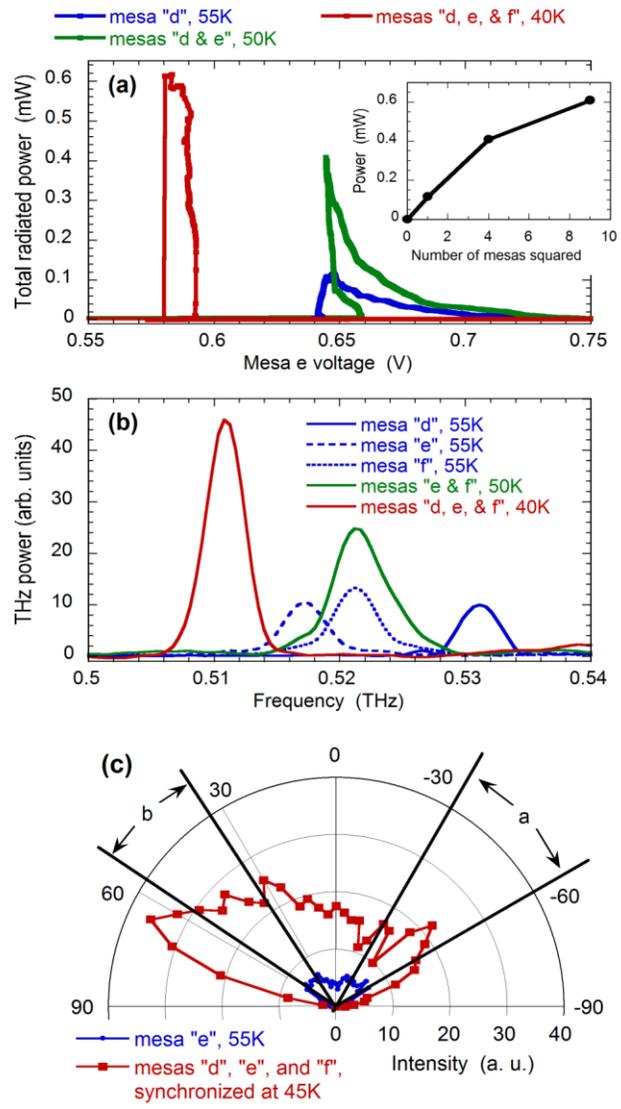

Figure 3

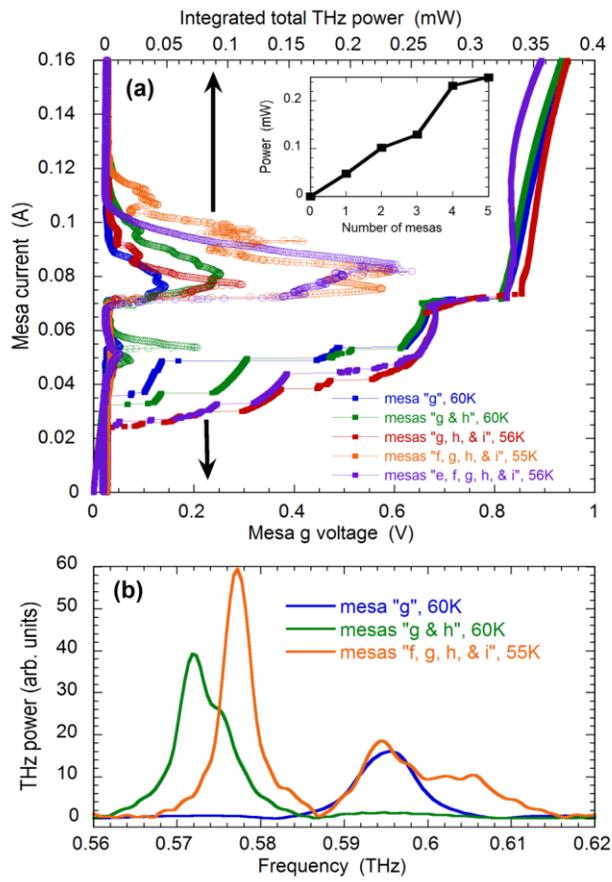

Figure 4

Supplemental Material for "**Powerful terahertz emission from Bi$_2$Sr$_2$CaCu$_2$O$_{8+d}$ mesa arrays**", T. M. Benseman et al.

Notes on THz measurement and detector calibration

Figure 1 below shows the optical set-up used in our measurements of the THz-emission from BSCCO-mesas. The intensity of THz radiation is measured with liquid helium-cooled silicon bolometer A. Simultaneously with total power measurements, Fourier transform spectroscopy can be performed with a Bruker 80v spectrometer, employing bolometer B as its detector. In order to obtain accurate power measurements, the optical responsivity of the bolometers was established, as well as the collection efficiency of the optical set-up comprised of lens, mirror and guide pipes.

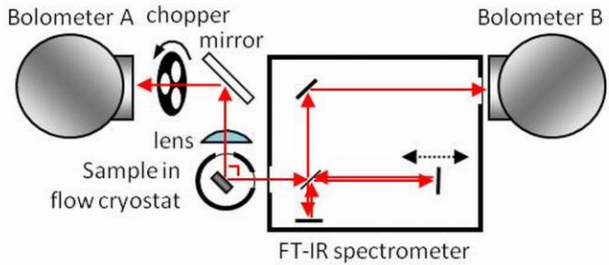

Supplemental Figure 1: Configuration of optical setup as used for mesa measurements, after reference [8].

The optical responsivity of the bolometers in the THz-frequency range was determined using the straight-shot set-up shown in Fig. 2 below. We use a blackbody source as power reference. The blackbody source employs multiple reflections from the surface of its cone-shaped interior walls to achieve an emissivity at the source aperture very close to 1, even if the cavity surface emissivity is less than 1 at the wavelength of interest. The aperture sizes and spacing are chosen such that all chopped radiation passing through the detection aperture is collected by the Winston cone. Note that the entrance diameter of the Winston cone is – in our case – half that of the exterior window of the bolometer. Then the detected power is given by the integral of the Planck function weighted by the filter and window transmission coefficients.

$$P_{detected} = \frac{\pi \, d_1^2 d_2^2}{16 \, l^2} \int_0^{\infty} \frac{C_1}{\lambda^5 [\exp(C_2/\lambda T) - 1]} T_{bandpass}(\lambda) T_{window}(\lambda) T_{long-pass}(\lambda) \, d\lambda$$

[Supplemental Equation 1]

where $C_1 = 3.74 \times 10^4$ W (μm)$^4$ cm$^{-2}$, and $C_2 = 1.44 \times 10^4$ W (μm)·K, $T$ is the cavity temperature in Kelvin, and $T_{bandpass}$, $T_{window}$, $T_{long\text{-}pass}$ are the various transmission coefficients.

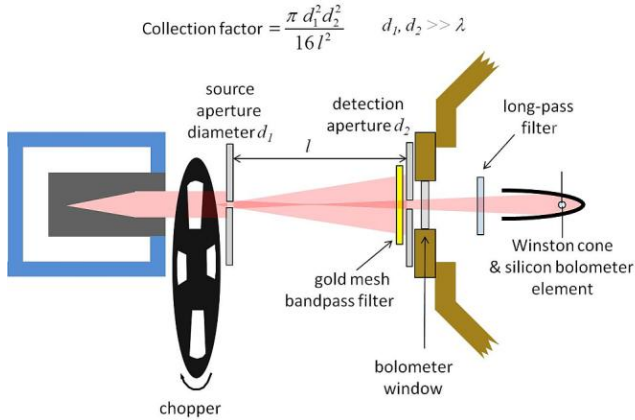

Supplemental Figure 2: Optical setup used for the calibration of the silicon bolometers.

Since the wavelengths transmitted by the filters (Supplemental Figure 3a) are all in the Rayleigh-Jeans limit at the cavity temperatures used here, the Planck formula can be reduced to

$$P_{blackbody}(\lambda) = \frac{C_1}{\lambda^4 C_2} T$$  [Supplemental Equation 2]

Then the integral in Supplemental Equation 1 scales linearly with the blackbody temperature, and the slope of the detector signal with respect to temperature (see Supplemental Figure 3b) gives the detector responsivity in Volts per Watt, at the centre wavelength of the bandpass filter.

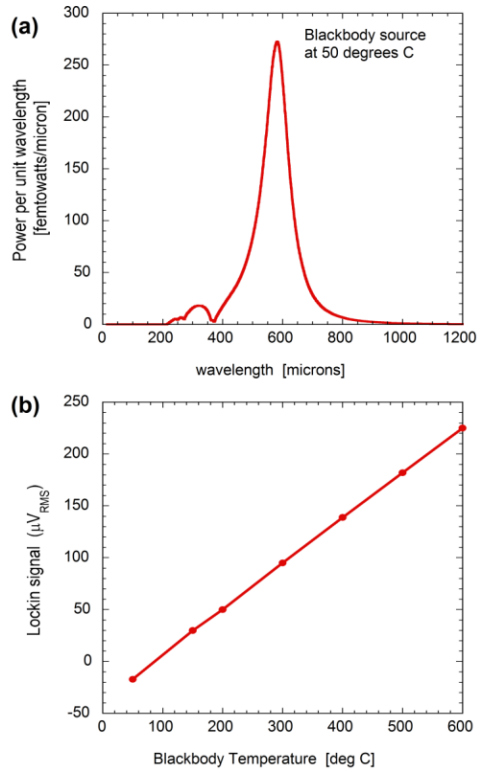

Supplemental Figure 3: (a) Example of the power spectrum delivered to the silicon bolometer, when using a gold mesh bandpass filter with center wavelength of 588 microns. (b) Bolometer output signal at 588 μm as a function of temperature, showing the expected linear *T*-dependence. The offset is due to the bolometer detecting the *difference* in radiated power from the blackbody source and the chopper blades.

The results of this procedure are shown in Figure 4 for bolometers A and B. We find that the optical responsivity of the detectors varies strongly with radiation wavelength, most likely due to wavelength-dependent absorbtivity of materials such as varnish used to bind the detector elements together. System # 1264 has an electrical responsivity of $3.07 \times 10^8$ V/W (at the preamplifier output, when set to a preamplifier gain setting of 1000) while system # 3324 has nominally identical design, and an electrical responsivity of $3.56 \times 10^8$ V/W (also at the preamplifier output).

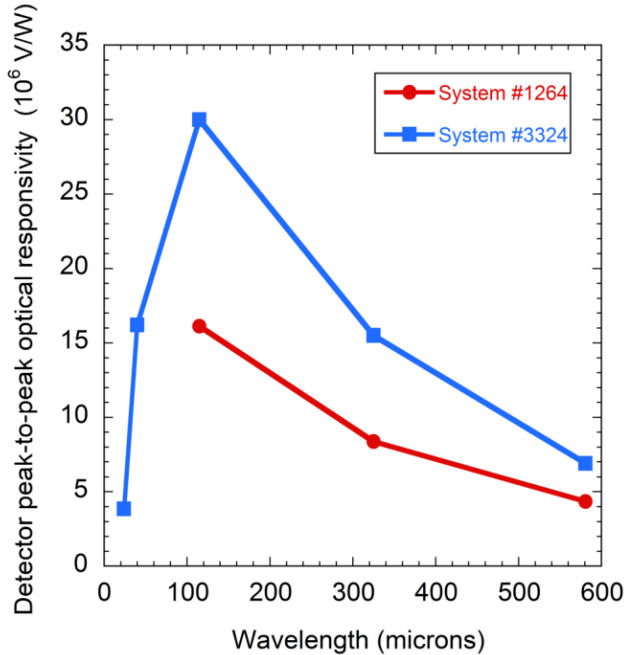

Supplemental Figure 4: Optical responsivity of the bolometers at various wavelengths, at preamplifier gain setting of 1000.

The collection efficiency of the optical set-up in Supplemental Fig. 1 was determined by measuring the emission power of a given mesa using the set-up as shown and then comparing to the result of a straight-shot measurement which has a known collection factor. In this way we determine a collection efficiency of 1.36 % for the set-up in Fig. 1. In this determination we included the angular dependence of the emission from the mesa.